\documentclass[%
 preprint,
superscriptaddress,
 amsmath,amssymb,
 aps,
prb,
floatfix,
]{revtex4-2}

\usepackage{graphicx}
\usepackage{dcolumn}
\usepackage{bm}

\usepackage{gensymb}

\makeatletter
\newcommand{\pushright}[1]{\ifmeasuring@#1\else\omit\hfill$\displaystyle#1$\fi\ignorespaces}
\newcommand{\pushleft}[1]{\ifmeasuring@#1\else\omit$\displaystyle#1$\hfill\fi\ignorespaces}
\makeatother

\makeatletter
\newif\ifpreprintoption
\@ifclasswith{revtex4-2}{preprint}{\preprintoptiontrue}{\preprintoptionfalse}
\makeatother
\newcommand{\ifpreprint}[2]{\ifpreprintoption #1\else #2\fi}

\usepackage{hyperref}

\begin{document}


\title{Distinguishing Carrier Transport and Interfacial Recombination at Perovskite-Transport Layer Interfaces Using Ultrafast Spectroscopy and Numerical Simulation}

\author{Edward Butler-Caddle}
\affiliation{Department of Physics, University of Warwick.}
\author{K.\ D.\ G.\ Imalka Jayawardena}
\affiliation{Advanced Technology Institute, University of Surrey.}


\author{Anjana Wijesekara}
\affiliation{Department of Physics, University of Warwick.}

\author{Rebecca L.\ Milot}
\affiliation{Department of Physics, University of Warwick.}

\author{James Lloyd-Hughes}
\email{j.lloyd-hughes@warwick.ac.uk}
\affiliation{Department of Physics, University of Warwick.}

\date{\today}

\begin{abstract}
In perovskite solar cells, photovoltaic action is created by charge transport layers (CTLs) either side of the light-absorbing metal halide perovskite semiconductor. 
Hence, the rates for desirable charge extraction and unwanted interfacial recombination at the perovskite-CTL interfaces play a critical role for device efficiency.
Here, the electrical properties of perovskite-CTL bilayer heterostructures are obtained using ultrafast THz and optical studies of the charge carrier dynamics after pulsed photoexcitation, combined with a physical model of charge carrier transport that includes the prominent Coulombic forces that arise after selective charge extraction into a CTL, and cross-interfacial recombination.
The charge extraction velocity at the interface and the ambipolar diffusion coefficient within the perovskite are determined from the experimental decay profiles for heterostructures with three of the highest performing CTLs, namely C$_{60}$, PCBM and Spiro-OMeTAD. 
Definitive targets for the further improvement of devices are deduced: fullerenes deliver fast electron extraction, but suffer from a large rate constant for cross-interface recombination or hole extraction. 
Conversely, Spiro-OMeTAD exhibits slow hole extraction but does not increase the perovskite's surface recombination rate, likely contributing to its success in solar cell devices.
\end{abstract}

\maketitle

\ifpreprint{\newpage}{}

\section{Introduction}

Perovskite solar cells first gained widespread attention when their power conversion efficiency jumped to over 10\% by changing from a dye-sensitised structure incorporating a liquid, to an all solid-state design \cite{Lee2012}. 
This structure remains the standard architecture to this day \cite{Jena2019,Jayawardena2020}, and consists of a light absorbing metal-halide perovskite sandwiched between an electron transporting layer (ETL) that selectively extracts conduction band electrons from the perovskite (but not valence band holes), and a hole transporting layer (HTL) that does the opposite, as pictured in Fig.\,\ref{fig:1-band diagrams}.
Employing interfaces of opposite selectivity on the two sides of the device gives photovoltaic action \cite{Wurfel2009,Wurfel2015}, which can be thought of as a kinetic asymmetry, and hence these heterojunctions are crucial to device performance. 
Selective extraction is enabled by a good electronic band alignment for one band at the interface (e.g.\ perovskite conduction band and ETL LUMO in Fig.\,\ref{fig:1-band diagrams}a), and the other band being misaligned (e.g.\ perovskite valence band and ETL HOMO). 
Photovoltaic action also arises from the differences in the equilibrium Fermi levels of the different materials before they are connected \cite{Wurfel2009,Wurfel2015} (particularly the electrodes), which creates a built-in field and an opposing gradient in chemical potential across the device when the layers are connected \cite{Hara2013,Lipovsek2019} (Fig.\,\ref{fig:1-band diagrams}b). Under illumination, the combined electrochemical potential develops a net gradient that drives electrons and holes towards different electrodes.  
Although photovoltaic action can occur without selective extraction layers, the charge-carrier blocking function of the charge transport layers (CTLs) has been found to be crucial for high performance, by preventing electron-hole recombination processes at the electrodes that compete with power extraction \cite{Juarez-Perez2014,Zhang2015,Liao2020}.
As the quality of the perovskite layer has improved (i.e.\ recombination rate constants within the perovskite bulk have been reduced), focus has shifted towards increasing the performance of CTLs \cite{Wolff2019,LeCorre2019,Yoo2022}. 
A wide variety of CTL materials have been trialled but those used in the earliest devices have remained amongst the most successful, namely the hole extracting material Spiro-OMeTAD, and the electron extracting materials C$_{60}$ (fullerene molecule) and its derivative PCBM. 

\begin{figure} [tb]
    \centering
    \includegraphics[width=0.4\textwidth]{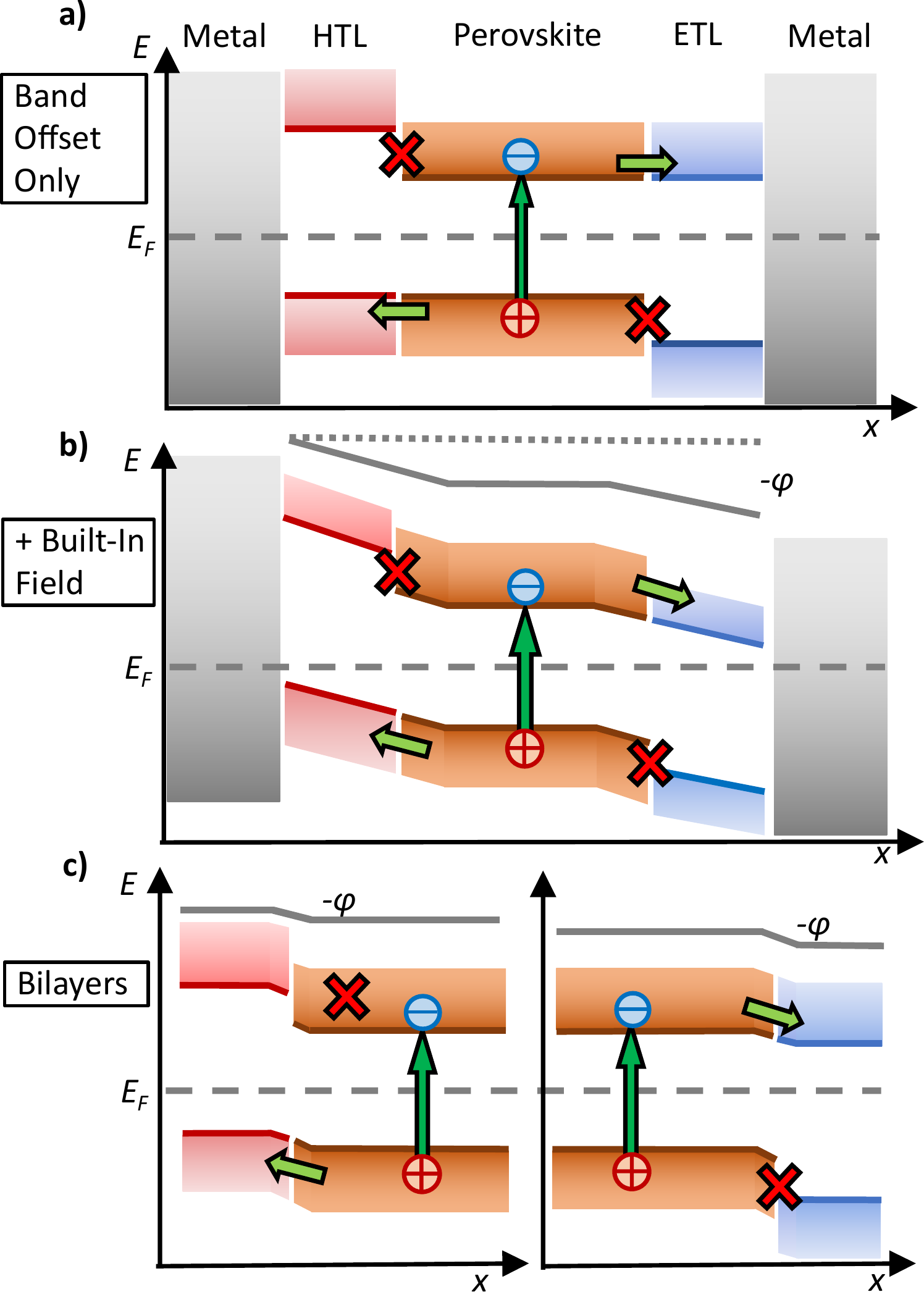}
    \caption{(a)-(b) Band diagrams for a typical perovskite solar cell geometry at equilibrium, with a hole transport layer (HTL) and electron transport layer (ETL) sandwiching the perovskite. (a) Kinetic asymmetry is induced by the band alignments of HTL and ETL relative to the perovskite. (b) The differences in Fermi level of each material creates built-in electric fields, mostly due to the electrodes. (c) Band diagrams for a perovskite-HTL bilayer (left) and a perovskite-ETL bilayer (right), with small built-in fields confined to thin layers near the interfaces.
    }
    \label{fig:1-band diagrams}
\end{figure}

In order to make further increases in performance by design, rather than slow trial and error, the electrical properties of the heterojunctions formed by these CTLs should be understood, so that the limiting factors can be targeted for improvement. 
Whilst measurements of the carrier dynamics of a complete device would be most representative of operational conditions, the large number of interfaces and processes present would make the dynamics difficult to interpret. 
Instead, by studying simple perovskite-CTL bilayers with one interface (for brevity shortened to CTL bilayers), the dynamics can be more easily interpreted and related to individual processes.
Although the built-in field in a bilayer will not be as large as in a full device (Fig.\,\ref{fig:1-band diagrams}b-c), properties of individual interfaces derived from bilayer measurements can be readily transferred to a model of the full solar cell. 
Ideally, to uncover the diffusion coefficient and the kinetics of interfacial processes, studies of the carrier dynamics of bilayers should have excellent time resolution, well below 1\,ns. 

Here we report an investigation of the interfacial kinetics for perovskite-CTL bilayers fabricated from three of the highest performing CTLs -- C$_{60}$, PCBM and Spiro-OMeTAD -- using time-resolved optical spectroscopy with photoluminescence, THz or visible white light probes.
We elucidate the carrier dynamics on sub-nanosecond timescales using a physical model that, critically, includes the Coulombic forces between electrons and holes.
This is vital for the description of interfacial kinetics when one charge carrier species is selectively extracted across an interface: for example this can generate Coulomb forces that act as a bottleneck, switching off charge extraction.
A quantitative agreement between the THz photoconductivity dynamics and the carrier dynamics model enabled the rate constants for surface extraction and cross-interfacial recombination to be determined, which revealed that C$_{60}$ and PCBM both rapidly extract electrons, but unfortunately have sufficiently fast cross-interface recombination that no Coulomb bottleneck forms.
In marked contrast, Spiro-OMeTAD did not rapidly extract holes from the perovskite, but did have a substantially slower surface extraction velocity and surface recombination velocity. 
The trends observed with sub-nanosecond THz and transient absorption measurements were also observed in nanosecond photoluminescence transients at lower injected carrier densities (comparable to solar illumination), indicating that the properties extracted from the sub-nanosecond measurements using higher injected carrier densities are relevant to operational carrier densities.

\section{Experimental Results}
\subsection{Heterojunction Materials}

For photovoltaics, the most directly relevant perovskite compositions to develop are those that have bandgaps close to the optimum calculated by the Shockley-Queisser model, whilst being sufficiently stable against degradation.
One of the most promising routes to rapidly producing cells that exceed the efficiency of silicon cells is to form perovskite-silicon tandem cells, for which the Shockley-Queisser model gives an optimum perovskite bandgap of 1.6-1.7\,eV, \cite{Aydin2020,DeBastiani2021} which requires mixed-halide compositions containing a majority of iodine and a minority of bromine.
In this work, spin-coated thin films (for fabrication details see Appendix) of triple cation mixed-halide perovskite were employed, with the composition $($FA$_{0.83}$MA$_{0.17)})_{0.95}$Cs$_{0.05}$Pb(I$_{0.83}$Br$_{0.17}$)$_3$. 
This composition has high crystallinity and excellent reproducibility and stability, including against light-induced halide segregation \cite{Saliba2016}. 
Electron micrographs (Fig.\,\ref{fig:2-SEM PL}\,a-c) showed complete coverage of the quartz substrates, with grain diameters of a few hundred nanometres, with most grains extending through the thickness of the layer (either 370\,nm or 600\,nm thick depending on the batch). 
\begin{figure*} [tb]
    \centering
    \includegraphics[width=0.95\textwidth]{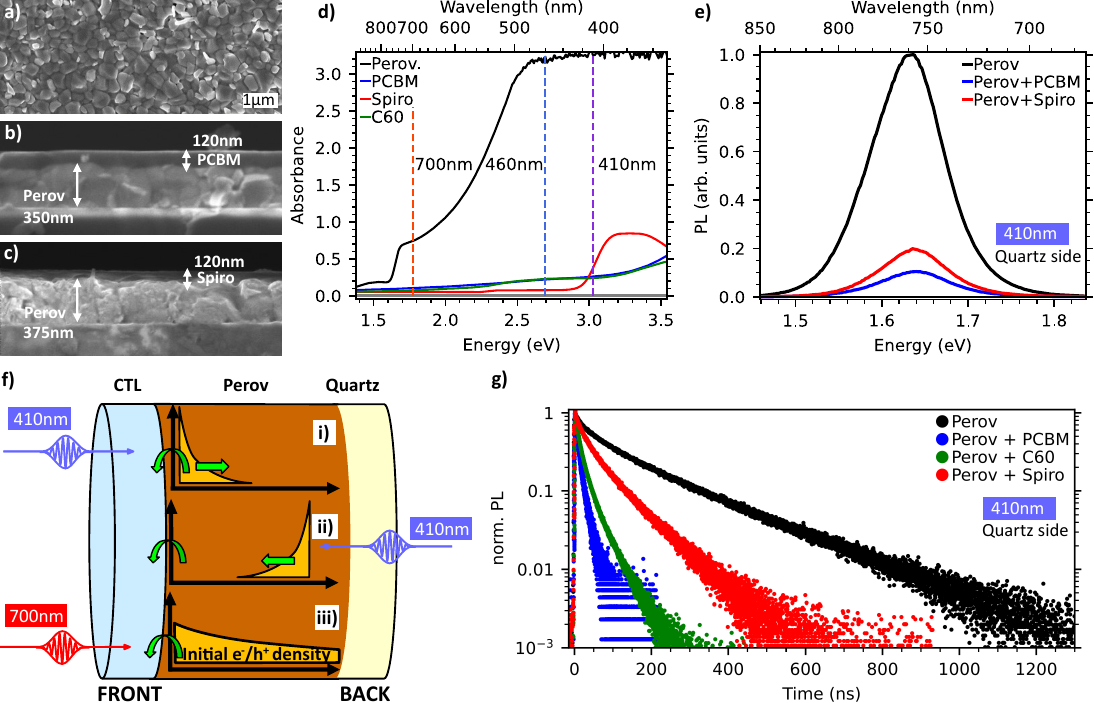}
    \caption{(a)-(c) Representative SEM images of perovskite/CTL bilayers, showing: (a) grain morphology of the front surface; (b)-(c) cross-sections of perovskite/PCBM and perovskite/Spiro OMeTAD bilayers. (d) Absorbance spectra of individual materials used. The absorbance of the perovskite film (370\,nm thick) at energies above 2.5\,eV exceeded the maximum absorbance measurable with the spectrometer, and hence was estimated from that for thinner films (see SM section 1.3 \cite{Note1}). (e) Photoluminescence (PL) spectra of samples (600\,nm thick) under 410\,nm CW excitation from the back (quartz) side ($\sim 400$\,Wm$^{-2}$). (f) Schematic of the different carrier profiles created with different pump wavelengths or excitation from alternate sides. (g) Time-resolved PL dynamics (at the peak PL energy) under 410\,nm pulsed excitation from the back (370\,nm thick).
    }
    \label{fig:2-SEM PL}
\end{figure*}

After forming the perovskite thin films, different CTLs were deposited on top to form various perovskite-CTL bilayers. 
In the bilayers with spin-coated PCBM or Spiro-OMeTAD layers, the CTLs were approximately 120\,nm thick, while thermally evaporated C$_{60}$ was 30\,nm thick (AFM measurements confirmed complete coverage). 
These layers were thicker than used in solar cells, in order to ensure complete coverage of the underlying perovskite and thus avoid adding another variable, \textit{i.e.}\ surface coverage, that can modify the dynamics (for example, with incomplete coverage some carriers must travel an additional distance to reach a perovskite-CTL heterojunction, and the carriers extracted into the CTL will be more concentrated). The ultrafast carrier dynamics should not be influenced by thicknesses above a few nanometres because within the 3\,ns experimental time window, the low mobilities of the CTLs mean that extracted carriers will travel less than a few nanometres in fullerenes, and even less in Spiro-OMeTAD. 
To check whether the solution deposition of PCBM and Spiro-OMeTAD affected the underlying perovskite, perovskite layers washed with chlorobenzene were also prepared and measured, and showed no significant difference (see Supplemental Material, SM, \footnote{See Supplemental Material [url] for: additional optical data; details of the numerical and analytical models; and additional simulation results. It includes additional Refs.\,\cite{
Karakus2015, Farrell2017, Evans2000, Vasileska2010, Eymard2000, LeVeque1990}.} Figs.\,S8 and S9). 
Separately, reference CTLs were deposited directly on quartz to confirm their optical absorbance.

The UV and visible absorbance of the individual materials is reported in Fig.\,\ref{fig:2-SEM PL}d, showing that while PCBM and C$_{60}$ exhibit a small amount of absorption across the visible, Spiro-OMeTAD has strong absorption above 3.0\,eV. 
The triple-cation perovskite had a sharp absorption onset (band edge) near 1.6\,eV and its steady state photoluminescence (SS-PL) spectrum (Fig.\,\ref{fig:2-SEM PL}e) peaked at 1.63\,eV, which demonstrated this composition's suitable bandgap energy for tandem solar cells. 
Furthermore, SS-PL spectra showed no evidence of light induced halide segregation as a second red-shifted peak did not appear under illumination, evidencing the long-term stability of this composition. The reduction in SS-PL intensities for the bilayers compared to the bare perovskite, has been interpreted as an increase in non-radiative recombination channels when the CTLs are connected \cite{Sarritzu2017, Stolterfoht2018, Hutter2020}.

Throughout this paper, measurements on bare perovskite layers are plotted in black/grey, perovskite-PCBM bilayers are plotted in blue, perovskite-C60 bilayers are plotted in green, and perovskite-Spiro-OMeTAD bilayers are plotted in red.

\subsection{Time-resolved PL spectroscopy}
In time-resolved optical spectroscopy a signal $S(t)$ related to the carrier density in the perovskite layer is measured as a function of time $t$ after photoexcitation, where $S$ refers to either the PL intensity, transient absorption change, or THz photoconductivity (from optical pump, THz probe, or OPTP, spectroscopy), depending on the technique used.
While in the reference perovskite material the carrier density can be altered only by bulk carrier recombination or surface recombination near the edges of the perovskite crystallites, in bilayers $n$ can be additionally lowered by charge extraction into the CTL or by extra recombination at the CTL-perovskite interface. 
The experimental $S(t)$ corresponds to the total change integrated over the whole charge carrier distribution, and information about the density profile $n(x)$ with depth $x$ is therefore lost from data at one time $t$.
However in certain situations the dynamical signal, $S(t)$, can still yield useful information about processes that occur at different depths in the bilayer, particularly if different photoexcitation schemes are used. 
As pictured in Fig.\,\ref{fig:2-SEM PL}f, illuminating with short wavelengths generates carriers in the perovskite near the CTL (case i), and hence the initial charge extraction and/or interfacial recombination rates should be high.
At later times these rates should drop, as charges diffuse away from the heterojunction into the perovskite.
In contrast, excitation through the substrate (case ii) yields carriers that are well separated from the CTL interface, with a low initial decay rate that then increases after carriers have had sufficient time to diffuse through the perovskite layer. 
Illuminating with long wavelengths (case iii) generates carriers throughout the depth of the perovskite layer, with similar dynamics expected to the long time limit for cases i and ii. 

Time-resolved PL dynamics measured by time-correlated single photon counting (TCSPC) are reported in Fig.\,\ref{fig:2-SEM PL}g for the perovskite-CTL bilayers and bare perovskite reference, recorded under 405\,nm excitation from the quartz side with incident pulse fluences of 25\,nJcm$^{-2}$ ($\sim 5\times 10^{14}$ incident photons m$^{-2}$ per pulse) and under ambient conditions in order to passivate surface defects (see Experimental Methods). 
The injected carrier density is comparable to that under solar illumination, which for this triple-cation perovskite composition has been estimated as $4 \times 10^{21}$\,m$^{-3}$ \cite{Hempel2022}. The modelling in Section \ref{Sec:modelling} indicates that carriers spread throughout the layer thickness within the few-nanosecond resolution of TCSPC, so an approximately uniform density can be considered, which gives an injected density of $10^{21}$\,m$^{-3}$ for our TCSPC measurements.

The bare perovskite (black curve) had a PL lifetime of $\tau_{\mathrm{eff}}=200$\,ns, consistent with results on a similar sample from electronically delayed OPTP (E-OPTP) with a long time window and low injection levels \cite{Butler-Caddle2023}, which represents the effective lifetime from bulk recombination and any surface recombination at the perovskite-air or perovskite-substrate interfaces.
When fullerene layers were added, the PL decay was significantly accelerated (blue and green curves), while the Spiro-OMeTAD bilayer (red) had only a marginally faster decay rate than the bare perovskite. 
The faster PL decay rates for the bilayers implies either charge extraction to the CTL or recombination at the CTL interface. The shortening of the PL decay time is consistent with the quenching of the steady-state PL intensity.

For each sample no difference in PL dynamics was observed for photoexcitation on opposite sides, despite generating different initial distributions due to the short penetration depth ($\delta \sim 35$\,nm) at this excitation wavelength. 
Furthermore, for each sample using excitation light with a longer penetration depth ($\delta \sim 260$\,nm for 633\,nm excitation) resulted in similar dynamics as for 405\,nm excitation (see SM section 1.5 \cite{Note1}). 
This implies that the instrument response time (several nanoseconds) was not sufficiently fast to distinguish case (i)(excitation on the front with short penetration depths), which should have much higher initial rates of extraction and interfacial recombination  compared to cases (ii) and (iii) (excitation on back or excitation with long penetration depths).
In short, the time taken for carriers to diffuse through the perovskite film is lower than the instrument response time in these PL experiments, which therefore probed the later dynamics only.

\subsection{Ultrafast spectroscopy}
In order to investigate charge transport, charge extraction and interfacial recombination, we therefore turned to ultrafast pump-probe spectroscopies with sub-picosecond time resolution. 
These were performed with incident pulse fluences 
of 2-7\,$\mu$Jcm$^{-2}$, corresponding to an incident photon flux of $0.6-1.4 \times 10^{17}$\,m$^{-2}$ per pulse.
The pump fluences for different wavelengths were chosen such that the absorbed photon fluxes per pulse were similar for the different pump wavelengths (using the absorption data reported in Supplemental Fig.\,S2 and S3 \cite{Note1}).
These low pump fluences were chosen such that the OPTP and TA decay curves were flat for the bare perovskite sample (see Fig.\,\ref{fig:3-OPTP} and Fig.\,\ref{fig:4-TA}), indicating that the fraction of the carrier population that recombines during the measurement window of 3\,ns is negligible. This means recombination within the perovskite can be excluded, which significantly simplifies the interpretation of the data. 
It is worth noting that the ultrafast dynamics were found to be more reproducible than time-resolved PL measurements performed at much lower pulse fluences of 25\,nJcm$^{-2}$ ($\sim 5\times 10^{14}$ incident photons m$^{-2}$ per pulse). 

\subsubsection{Origin of the transient photoconductance}
The OPTP decays reported in Fig.\,\ref{fig:3-OPTP} show the change in transmitted electric field, $S(t)=-\Delta T/T$, which is proportional to the sheet photoconductance $\Delta\sigma$ of the sample \cite{Butler-Caddle2023}.
As the conductivity is $\sigma=n e\mu_e + p e \mu_h$, OPTP is sensitive to changes in either mobility or density, and hence we first establish the origin of $S(t)$ for CTL-perovskite bilayers.
No appreciable contribution to the photoconductivity is expected from charges transferred to the CTLs, where carriers have a substantially lower mobility than in the perovskite: THz sum mobilities $\mu_e+\mu_h$ exceed 10\,cm$^{2}$Vs$^{-1}$ in perovskites \cite{Hempel2022}, whereas Spiro-OMeTAD has $\mu_h<10^{-3}$\,cm$^{2}$Vs$^{-1}$ \cite{Shi2016} and PCBM exhibits $\mu_e<1$\,cm$^{2}$Vs$^{-1}$ \cite{Priebe1997,Cabanillas-Gonzalez2006,Singh2007}.

To validate this assumption, in this work CTLs deposited on quartz were photoexcited above their bandgap (at 410\,nm), and no CTL photoconductivity was observed via OPTP at the experimental fluences used. 
Hence we conclude that $S(t)$ arises just from carriers in the perovskite film.
Within the first picosecond following photoexcitation, the mobility of hot photoexcited carriers in the perovskite may change as carriers relax within the bandstructure \cite{Monti2018,Monti2020} or into localised states \cite{Hempel2022}.
However after this time the mobility can be assumed to be limited by electron-phonon scattering, and independent of time (and density), and thus changes in $-\Delta T/T$ are linked to the density of photoexcited carriers in the perovskite. 

The sum of the electron and hole mobilities was calculated for the perovskite reference sample from the initial amplitude of the OPTP signal (method described in \cite{Ren2023}), assuming unity quantum yield of free carriers (reasonable due to the low exciton binding energy in these perovskites \cite{Baranowski2020}).
This yielded a sum mobility $\sim 40$\,cm$^2$V$^{-1}$s$^{-1}$ when using short pump wavelengths (and $\sim 70$\,cm$^2$V$^{-1}$s$^{-1}$ when using long pump wavelengths), consistent with literature \cite{Hempel2022}.
Hence the detected $S(t)$ from OPTP for bilayers can be safely assumed to arise only from the photoconductance of mobile charges in the perovskite films.

\begin{figure*}[tb]
    \centering
    \includegraphics[width=0.85\textwidth]{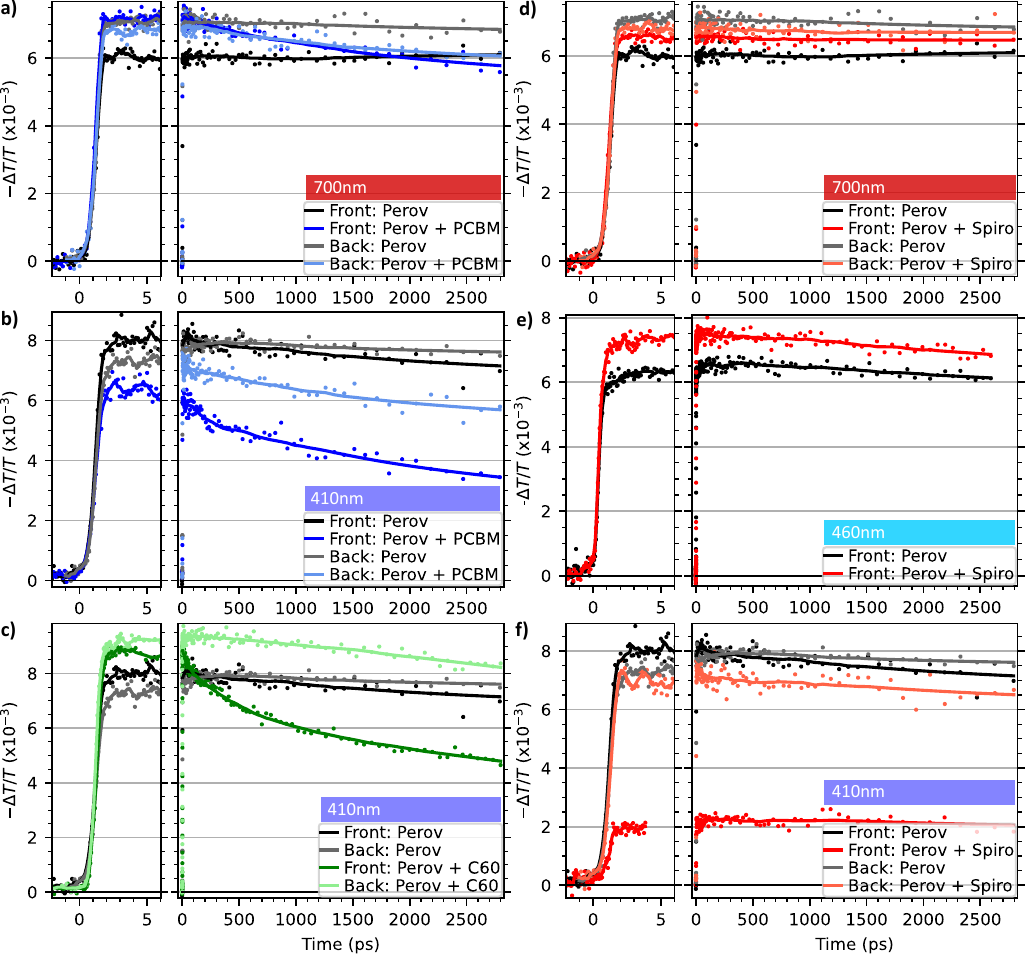}
    \caption{(a)-(c) OPTP signal $S(t)=-\Delta T/T$ for perovskite-CTL bilayers with fullerene ETLs (C60 or PCBM) under different excitation wavelengths (all with incident pulse fluences of 2-7\,$\mu$Jcm$^{-2}$, or 0.6-1.4$\times 10^{17}$ photons m$^{-2}$ per pulse). Points show the raw data while the solid lines show a running average. Data for the bare perovskite reference (washed with chlorobenzene) are shown in black (front excitation) and grey (back excitation). Perovskite-PCBM bilayers are shown in blue, while C$_{60}$ is in green, where darker and lighter shades indicate front and back excitation, respectively. (d)-(f) As (a)-(c), but for bilayers with the hole transporter Spiro OMeTAD. All samples had 600\,nm thick perovskite, except e) was 370\,nm thick.}
    \label{fig:3-OPTP}
\end{figure*}

\subsubsection{Transient photoconductance of perovskite-ETL bilayers}

Fig.\,\ref{fig:3-OPTP}(a-c) reports $S(t)=-\Delta T/T$ for the fullerene bilayers (C$_{60}$ in green and PCBM in blue), which exhibited accelerated decay over the 3\,ns window relative to the bare perovskite sample (black/grey), for both ``front'' excitation through the CTL (darker shade), and ``back'' excitation through the substrate (lighter shade). 

For 700\,nm excitation on the PCBM bilayer (case iii, Fig\,3a), the initial amplitude and the decay dynamics for front excitation (dark blue) were similar to those for back excitation (light blue). 
This is consistent with the negligible parasitic absorption of this pump wavelength by the PCBM (Fig.\,\ref{fig:2-SEM PL}d), and with the expectation that carrier removal dynamics will be similar when the electrons and holes are photoinjected throughout the layer (the perovskite's absorption depth was $\sim 260$\,nm). 
For excitation with 410\,nm (Figs.\,3b-c) on the front of the fullerene bilayers (case i), the decay was markedly faster than for 700\,nm excitation: the carriers were concentrated near the interface rather than spread throughout the perovskite film's thickness. 
The initial amplitude of the signal was marginally lower for excitation through the front (case (i), dark blue and green) compared to through the back (case (ii), light blue and green), in agreement with parasitic absorption of the pump by the fullerenes (Fig.\,\ref{fig:2-SEM PL}d).

For 410\,nm excitation on the back (case (ii)) of the C$_{60}$ bilayer (light green, Fig.\,\ref{fig:3-OPTP}c), the decay rate was slow at early pump-probe delay times, before accelerating at later times. 
This can be qualitatively understood as requiring a finite time for carriers to diffuse from a low recombination velocity surface (perovskite-quartz interface) through the thickness of the film to the CTL interface, where population reduction can then occur. 
A quantitative model for both front and back excitation is discussed in Section 3.1. 
On the other hand, for 410\,nm excitation on the back (case (ii)) of the PCBM bilayer (light blue, Fig.\,\ref{fig:3-OPTP}b), the dynamics were similar to that for 700\,nm excitation. 
This suggests that the solution-deposited PCBM had penetrated deeper into the perovskite than the vapour deposited C$_{60}$, thus reducing the distance carriers must diffuse from the quartz side to reach the PCBM.

\subsubsection{Transient photoconductance of perovskite-HTL bilayer}
The corresponding OPTP data for the Spiro-OMeTAD bilayer (hole extracting) are reported in Fig.\,\ref{fig:3-OPTP}d-f. 
In stark contrast to the case of electron extraction, there is no faster decay in the OPTP signal in comparison to the perovskite reference over the 3\,ns window, when excited on either side of the sample or with any of the pump wavelengths used (410\,nm, 460\,nm and 700\,nm). 
The only substantial difference is in the initial amplitude when exciting with 410\,nm through the CTL: the Spiro-OMeTAD bilayer exhibited around 25\,\% of the photoconductivity of the perovskite reference.
In this case, a large fraction of the optical pump beam was absorbed by the Spiro-OMeTAD layer first, as Spiro-OMeTAD exhibits a substantial absorbance for that pump photon energy (Fig.\,\ref{fig:2-SEM PL}d; 410\,nm corresponds to 3.02\,eV). 
For excitation conditions that did not exhibit parasitic absorption -- 700\,nm and 460\,nm excitation from either side and 410\,nm back excitation -- the similar initial amplitude to that of the bare perovskite, and the lack of decay within the 3\,ns window, suggests that there is no substantial removal of carrier density at the Spiro-OMeTAD-perovskite interface within 3\,ns via selective charge extraction or via any additional surface recombination. 
This conclusion can be drawn because in all cases carriers were either generated next to the Spiro-OMeTAD interface, or would have diffused to the interface within the 3\,ns window, as evidenced by the fullerene measurements. 
Thus we have established that the rates for extraction and surface recombination at the Spiro-OMeTAD-perovskite interface must be low, such that the effective lifetime is close to the bulk lifetime. 
This is in agreement with the lifetimes from time-resolved PL, discussed further in Section 3.4.

\subsubsection{Transient absorption of perovskite-CTL bilayers}
An additional factor complicating the interpretation of OPTP spectroscopy of selective carrier extraction, is that $S(t)$ for OPTP is proportional to the sum of the photoconductivity of electrons and holes, and their carrier mobilities may differ.
Therefore, extraction of one carrier type may cause a different reduction in $S(t)$ than removal of the other type. 
Computational studies have suggested similar effective masses for both electrons and holes \cite{Giorgi2013}, but the ratio of their mobilities is harder to determine as mobility depends on the scattering rate as well. 
Therefore, to check that the lack of decay over 3\,ns for the Spiro-OMeTAD bilayers was not due to OPTP being less sensitive to hole density (if it were the case that $\mu_h\ll \mu_e$), white light TA spectroscopy was additionally performed, as reported in Fig.\,\ref{fig:4-TA}. 
In TA spectroscopy $S(t)=\Delta \mathrm{OD}$, the change in optical density.
The strength of the ground state bleach (negative $\Delta \mathrm{OD}$, red in Fig.\,\ref{fig:4-TA}a) of the interband transition can be assumed to be proportional to the carrier density injected, independently of the mobility.
Here, experiments were performed at low enough pump fluences that the hot phonon bottleneck regime \cite{Monti2020} was not reached, and carriers cooled quickly (within a few ps) to the band edge.
The TA dynamics (Fig.\,\ref{fig:4-TA}b) show the same trends as for OPTP, with no decay for the Spiro-OMeTAD bilayers over 3\,ns, and the fullerene bilayers did exhibit additional decay in comparison to the bare perovskite reference (the amplitudes are not interpreted since they were not as well controlled).
The agreement between TA and OPTP transients additionally confirms that the decay of the photoconductance signal is due to carrier removal rather than changes in mobility. 
\begin{figure*}[tb]
    \centering
    \includegraphics[width=0.95\textwidth]{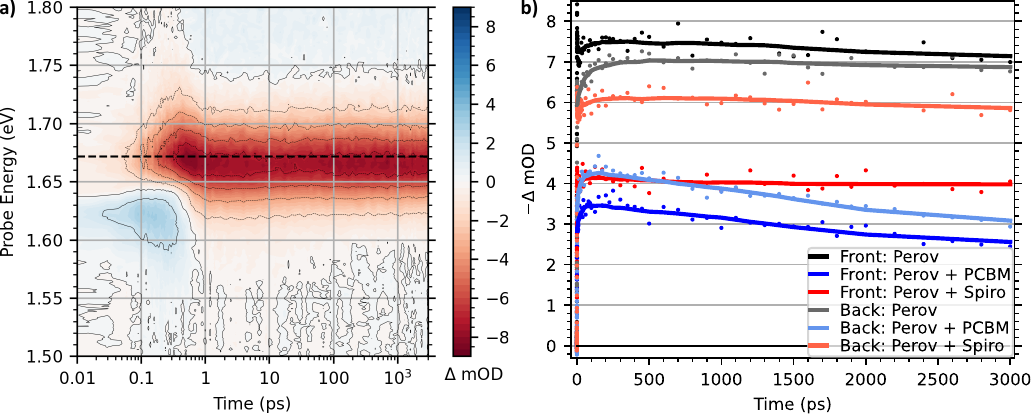}
    \caption{(a) Transient absorption spectra $\Delta$mOD versus probe energy and time delay for the perovskite reference (600\,nm thick) with 410\,nm excitation on the front (all with incident pulse fluences of 3\,$\mu$Jcm$^{-2}$, or $7\times 10^{16}$ photons m$^{-2}$ per pulse). The peak of the GSB is illustrated by the dashed line. 
    (b) Transient absorption dynamics at an energy of 1.67\,eV (the GSB for the perovskite). Darker and lighter shades indicate front and back excitation respectively: black = perovskite reference (washed with chlorobenzene), blue = perovskite-PCBM bilayer, red = perovskite-Spiro-OMeTAD bilayer. 
    }
    \label{fig:4-TA}
\end{figure*}

\section{Carrier dynamics modelling}\label{Sec:modelling}
We developed a theoretical model of the evolution of the charge density within a CTL-perovksite bilayer in order to simulate the dynamical signal $S(t)$ observed in experiments, and to disentangle the different contributions to $S(t)$ from diffusion within the perovskite, population transfer to the CTL and recombination at the perovskite-CTL interface. 
A selective interface will clearly lead to the spatial separation of electrons and holes, and the resulting Coulombic forces will influence the motion of carriers within the perovskite, \emph{i.e.}\ introducing drift transport in addition to diffusive motion. 
A mathematical description of $n(x,t)$ and $p(x,t)$ requires solving the non-linear continuity equations for electrons and holes, including drift and diffusion current terms, along with the electric field $E(x,t)$ obtained from Gauss's law.
Krogmeier \emph{et al.}\ \cite{Krogmeier2018} simulated the carrier dynamics of perovskite-CTL bilayers and suggested the importance of Coulombic forces at high carrier densities, such as used in the OPTP and TA experiments reported here.
They focused on modelling the dynamics measured by TRPL over tens of nanoseconds; here we instead focus on carrier dynamics on the sub-nanosecond timescales relevant to OPTP and TA measurements. 
Although device simulation packages are available \cite{Courtier2019,Calado2022}, these are not designed for simulating the transients of bilayers on picosecond timescales, following pulsed illumination. 

\subsection{Numerical model}
As described in the Methods and Supplemental Material \cite{Note1},
in the present work we developed a numerical solution using the Scharfetter-Gummel discretisation of the drift-diffusion current equations and the forward Euler method \cite{Selberherr1984}, for parameters that match our experimental conditions.
Perovskites are at most weakly doped \cite{Pena-Camargo2022}, and hence at the fluences used in OPTP and TA spectroscopy only the photoexcited carriers ($n_e, p_e$) are considered in the model, as these will have higher density than the equilibrium carriers ($n_0, p_0$), i.e.\ high level injection $n_e, p_e >> n_0, p_0$.
Experiments were performed at low pump fluences ($0.6-1.4\times 10^{17}$ photons m$^{-2}$ per pulse) such that the OPTP and TA decay curves were flat for the bare perovskite sample (see Fig.\,\ref{fig:3-OPTP} and Fig.\,\ref{fig:4-TA}), indicating that the fraction of the carrier population that recombined during the measurement window of 3\,ns was negligible. 
Thus recombination within the perovskite can be excluded, which significantly simplifies the interpretation of the data and reduces the number of free parameters in the model.
Further, the low fluence allowed the influence of photon reabsorption \cite{Crothers2017, Hutchinson2023} to be ignored in our model as the rate of bimolecular radiative recombination was minimised. The migration of ions is too slow to occur during the measurement window of a few nanoseconds so can be ignored.
\begin{figure}[t]
    \centering
    \includegraphics[width=0.9\textwidth]{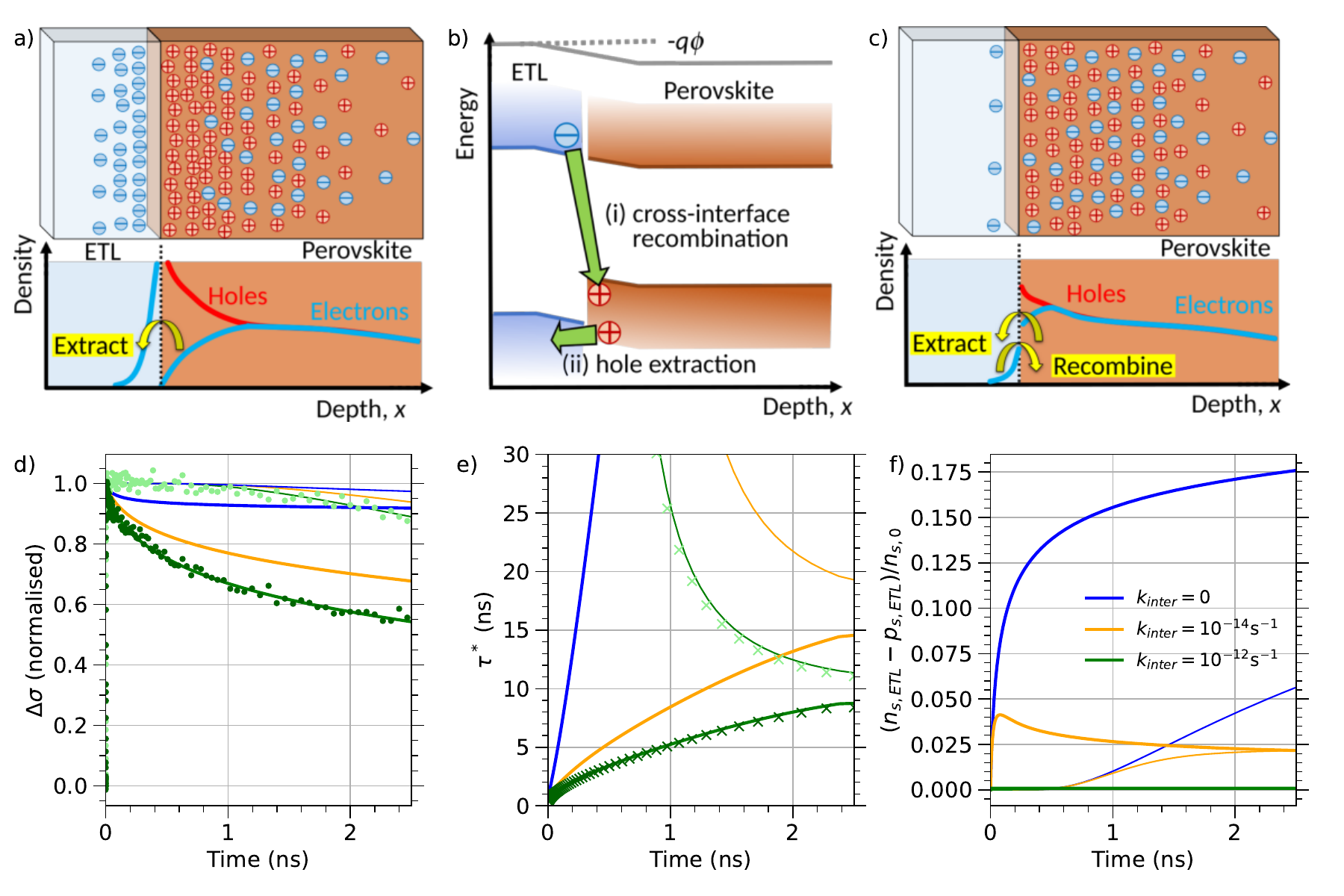}
    \caption{(a) Schematic of charge distribution in a perovskite-ETL bilayer around 100\,ps after photoexcitation through the ETL, without interface recombination. Electrons (blue) have been extracted into the ETL (grey), and an excess of holes (red) has built up in the perovskite near the interface, forming an electric double layer. (b) Schematic band diagram at the interface showing processes that lower the net charge in the ETL, either (i) an electron in the ETL recombining with a hole in the perovskite, or (ii) a hole moving from the perovskite to the ETL. (c) With strong interfacial recombination the electron density in the CTL is lower than that in panel (a), and the net charge difference in the perovskite near the interface is smaller. 
    (d) Simulated transient photoconductance $\Delta \sigma$ when photoexcited on the front (thick lines) or back (thin lines), either without interfacial recombination (blue), or with $k_{\mathrm{inter}}=10^{-14}$\,s$^{-1}$ (orange) or $k_{\mathrm{inter}}=10^{-12}$\,s$^{-1}$ (green), for an extraction velocity $S=90$\,ms$^{-1}$. 
    Points show the experimental data from Fig.\,\ref{fig:3-OPTP}c, for the perovskite-C$_{60}$ bilayer.
    (e) Effective lifetime $\tau^*(t) = -\Delta\sigma/[d(\Delta\sigma)/dt]$ calculated for the curves shown in (d), indicating that the decay rate changes dynamically with pump-probe delay. Crosses indicate results from the analytical model with $S=90$\,ms$^{-1}$ and $\mu=15$\,cm$^2$V$^{-1}$s$^{-1}$.
    (f) The excess sheet charge density in the ETL, $n_{s,ETL}-p_{s,ETL}$, shown normalised by the initial sheet charge density in the perovskite $n_{s,0}=1\times10^{13}$\,cm$^{-2}$.
    }
    \label{fig:5-Model}
\end{figure}

The physical processes involved in the simulation, are described in Fig.\,\ref{fig:5-Model}a-c for the case of electron removal via population transfer to the ETL, under excitation with an initial sheet carrier density of $10^{17}$\,m$^{-2}$ and a penetration depth of 35\,nm. 
Initially, the electrons and holes have the same profile in the perovskite (schematically shown in Fig.\,\ref{fig:5-Model}a), while at later times charges can either diffuse away from the interface into the perovskite, or be extracted to the ETL.
For a selective ETL we assumed that interfacial electrons were extracted at a rate defined by the interfacial extraction velocity, $S$, while hole extraction was forbidden.
The preferential extraction of electrons into the ETL (Fig.\,\ref{fig:5-Model}b) attracts holes in the perovskite towards the interface, and repels electrons, forming an electrical double layer. 
Further, in the simulation we considered two processes that can reduce the net charge in the ETL (Fig.\,\ref{fig:5-Model}c): (i) cross-interface recombination, where an electron in the LUMO of the ETL recombines with a hole in the VB of the perovskite at a rate $k_{\mathrm{inter}}n_{\mathrm{CTL}}p(x=0)$, or (ii), hole transfer from the perovskite VB to the HOMO of the ETL with velocity $S_{holes}$. 
This later process may be possible given the tail in the density of states for fullerene molecules above their HOMO level, suggesting that the VB offset between fullerene and perovskite is not very large and thus hole extraction may not be blocked very effectively \cite{Schulz2015}.
Cross-interface recombination and hole extraction are difficult to distinguish, as both lead to the loss of a hole from the perovskite and reduce the negative charge in the ETL (they are compared in SM section 2.5 \cite{Note1}).

The results of the simulations are shown in Figures 5d-f. As well as plotting the perovskite photoconductance versus pump-probe delay, the decay rate can be conveniently parameterised by the instantaneous lifetime, $\tau^*(t) = -\Delta\sigma/[d(\Delta\sigma)/dt]$. It should be noted that this represents the decay lifetime for the perovskite's entire sheet carrier density and is thus effected by the distribution of carriers. For example, even if the rate constant for carrier loss at an interface is fixed, the decay rate of the sheet density is faster if carriers are concentrated near the interface, and lower if they are spread throughout the thickness (the green lines, which will be discussed below, are an example of this).  

Fig.\,\ref{fig:5-Model}d shows simulations of the photoconductance $\Delta\sigma$, for an extraction velocity $S=90$\,ms$^{-1}$ and for various cross-interfacial recombination rates, under either front excitation (thick lines) or back excitation (thin lines).
Considering first the case without cross-interfacial recombination, $k_{\mathrm{inter}}=0$ (blue lines), for front excitation the simulated photoconductance of the perovskite initially decays before plateauing (thick line), while for back excitation the photoconductance only appreciably decays after about 1ns (thin line). 
The plateaux in the photoconductance corresponds to a rapidly increasing $\tau^*$ (blue line, Fig.\,\ref{fig:5-Model}e), and can be understood as resulting from the formation of the electrical double layer: the Coulomb repulsion of electrons away from the interface prevents further electron extraction, leading to the population decay rate slowing.
The electrical double layer's formation can be seen from the excess electron sheet density in the CTL, as reported in Fig.\,\ref{fig:5-Model}f.
The electron sheet density in the CTL can be seen to rapidly rise within 200\,ps to above 10\,\% of the initial sheet carrier density in the perovskite, before rolling over as the extraction rate slows. 

In line with the findings of Ref.\ \cite{Krogmeier2018}, simulations for carrier sheet densities three orders of magnitude lower (10$^{14}$ rather than 10$^{17}$ m$^{-2}$) do not show such a Coulombic bottleneck for extraction (see SM Fig.\,S22 \cite{Note1}).
To verify the robustness of the observed plateau in the simulated transients with $k_{\mathrm{inter}}=0$ (blue lines), we performed alternative simulations whereby electrons were allowed to be extracted from deeper into the perovskite layer (smearing the interface), or the perovskite's permittivity was increased to reduce the electric field strength (see SM Figs.\,S20 and S21 \cite{Note1}). 
However, even extreme smearing of the interface or extreme permittivity values were not sufficient to remove the plateau.

The simulations with $k_{\mathrm{inter}}=0$ described above are qualitatively different from the experimental $S(t)$ observed for the fullerene bilayers: the experiments did not show a plateau after a rapid decay; rather the OPTP, TA and TRPL exhibited a continual decay. 
Hence we can conclude that there is a physical process that prevents the formation of the Coulomb bottleneck for extraction, such as cross-interface recombination or hole extraction.
When cross-interface recombination was included by setting $k_{\mathrm{inter}}=10^{-14}$\,s$^{-1}$ (orange lines in Fig.\,\ref{fig:5-Model}d-f) the photoconductance continually decreased with time, and the extraction rate does not slow significantly.
This is a result of the cross-interface recombination reducing the strength of the Coulomb bottleneck: the excess sheet density in the ETL initially increases (Fig.\,\ref{fig:5-Model}f, orange line), but the accumulation of both $p(x=0)$ and the electron density in the ETL causes the cross-interfacial recombination rate to increase until it balances the extraction rate, meaning the net charge in the ETL stops rising and the Coulomb bottleneck is weak.
When the cross-interfacial recombination rate balances the extraction rate, extracted electrons can be considered to immediately undergo recombination.
With an even higher rate $k_{\mathrm{inter}}=10^{-12}$\,s$^{-1}$ no substantial charge density builds up in the ETL at all (green lines in Fig.\,\ref{fig:5-Model}f).
In this case, $\tau^*$ increases slowly due to carriers diffusing away from the interface (alluded to earlier), rather than the Coulombic bottleneck. Fig.\,\ref{fig:5-Model}d shows that parameter values of $S=90$\,ms$^{-1}$ and $k_{\mathrm{inter}}=10^{-12}$\,s$^{-1}$ give excellent agreement with the experimental data.

Summarising the results from the numerical simulation, the relative amplitude of the extraction and cross-interfacial recombination rate constants can be qualitatively obtained by visually inspecting the shape of the experimental decay at delay times below a few nanoseconds.
For Spiro-OMeTAD, population removal (extraction to the HTL or recombination) was not observed on timescales of 1\,ps to 3\,ns, and hence the extraction velocity $S$ and surface recombination are small.
For the fullerenes, the decay of the population within 3\,ns indicates that the charge extraction rate was higher, and the persistence of the decay over 3ns indicates that interfacial removal/loss processes, such as cross-interface recombination or hole extraction, limit the formation of a Coulomb bottleneck. 
In the following sections we quantify the extraction velocity $S$ and $k_{\mathrm{inter}}$.

\subsection{Analytical ``one-step'' model}
In the limiting case of rapid cross-interfacial recombination or simultaneous extraction of both carriers (such as for fullerene bilayers), we now discuss a simpler theoretical model of the carrier dynamics with fewer free parameters than the numerical model.
With rapid cross-interfacial recombination, the two-step process of extraction then recombination is effectively a one-step process limited by the slowest step -- extraction.
Therefore, this effective surface recombination has a velocity given by the extraction velocity.
This process means the electron and hole densities at the interface remain similar, and the carrier removal rate becomes proportional to the electron density in the perovskite at the interface.
This scenario satisfies the neutrality approximation, $n(x)=p(x)$, so there is no need to solve Gauss's law and the two continuity equations can be combined into a single equation for the balanced density $n(x)=p(x)$ \cite{McKelvey1966}. 
For high level injection, this ambipolar continuity equation is just the diffusion equation, and if the effective surface recombination is linear in density, then the equation can be solved analytically by separation of variables \cite{Hahn2012}:
\begin{equation}
n(x,t) = e^{-t/\tau_B} \sum_{m=1}^{\infty} A_m \cos(\alpha_m(L - x)) e^{- \alpha_m^2 D t}
\label{Eq:Fourier}
\end{equation}
\noindent where $\tau_B$ is the bulk recombination time in the perovskite and $\alpha_m$ are the different eigenvalue solutions of the transcendental equation $\alpha_m \tan(\alpha_m L) = S/D$  (see SM section 2.3 \cite{Note1}). 
Hence $\alpha_m$ depend on $L$, $S$ and $D$, where the surface recombination velocity $S$ is equal to the electron extraction velocity used in the numerical model, and $D$ is the \emph{ambipolar} diffusion coefficient.
The Fourier coefficients, $A_m$, are determined by the initial distribution and therefore by the absorption coefficient at the pump wavelength (from UV-visible spectroscopy).
Similar expressions have been used to analyse surface recombination in silicon \cite{Luke1987, Sproul1994, Butler-Caddle2023}, where $D$ and $L$ are substantially larger than in perovskites.
In order to compare the simulated sheet densities (an even simpler Fourier Series than Eq.\,\ref{Eq:Fourier}) to the measured $S(t)=-\Delta T/T$ we used the standard thin film expression \cite{Burdanova2019}.
Overall, the only free parameters in the model were the ambipolar diffusion coefficient, $D$, and the interfacial extraction velocity, $S$. 
Using this expression with $S=90$\,ms$^{-1}$ and $\mu=15$\,cm$^2$V$^{-1}$s$^{-1}$ (or $D=0.39$\,cm$^2$s$^{-1}$) yields $\tau^*(t)$ shown by the green points reported in Fig.\,\ref{fig:5-Model}e, which are almost identical to the numerical simulation's results (solid green lines) for the same parameters.

At late times, the instantaneous lifetime tended to a constant, namely $\tau^*(t\rightarrow\infty)\rightarrow 10$\,ns in the case shown in Fig.\,\ref{fig:5-Model}e, and was similar for both front and back excitation.
In this limit of long time delays, the carrier distribution described by Eq.\,\ref{Eq:Fourier} is dominated by the smallest eigenvalue $\alpha_m$, because it has the slowest temporal decay (and also has the smoothest spatial distribution). 
The distribution becomes stationary: the shape doesn't change, only the amplitude alters, which decays with rate  $1/\tau_{eff} = 1/\tau_{B} + 1/\tau_{S}$, where $1/\tau_{S} = \alpha_0^2D$ and $\tau_B$ is the bulk recombination lifetime. 
For a semiconductor with large $S$ on one surface (the extraction layer here), and negligible $S$ on the other surface, $\tau_S$ can be approximated as \cite{Sproul1994}:
\begin{equation}
    \tau_s = \frac{L}{S} + \frac{4}{D}\left(\frac{L}{\pi}\right)^2.
    \label{Eq:TauS}
\end{equation}
\noindent In contrast to representing surface recombination, as witnessed in a single semiconductor \cite{Sproul1994}, here for the perovskite-CTL bilayer $S$ represents extraction at speed $S$ followed by fast cross-interface recombination.
The second term in $\tau_s$ accounts for the time required for charges to diffuse through the film's thickness before reaching the interface.
Calculating the surface lifetime using Eq.\,\ref{Eq:TauS} for the same parameters as the numerical and analytical simulations in Fig.\,\ref{fig:5-Model} (i.e.\ $L=600$\,nm, $S=90$\,ms$^{-1}$, $D=0.39$\,cm$^2$s$^{-1}$) yields $\tau_s=10.5$\,ns and thus $\tau_{\mathrm{eff}}=10.0$\,ns, in excellent agreement with $\tau^*(t\rightarrow\infty)$ deduced from the models.

\subsection{Global fits to experiment}
As an analytical expression, Eq.\,\ref{Eq:Fourier} can be fit iteratively to the experimental data orders of magnitude faster than the full numerical solution. 
By globally fitting a set of decay curves for different experimental conditions, using a common set of parameters, one can obtain a sharper minimum of the global cost function with respect to the free parameters than is possible for a fit to just a single decay curve. 
Here, including front and back excitation with a short penetration depth gave a strong constraint on the ambipolar $D$ (and mobility), as the shape of the decay is strongly influenced by the time taken for carriers to diffuse through the perovskite to the CTL interface.
It is important to note that because the ambipolar continuity equation is linear, the normalised decay shape is the same regardless of the sheet density, and any error in the estimation of the initial sheet density has no impact on the parameters deduced. 
\begin{figure*}[tb]
    \centering
    \includegraphics[width=0.95\textwidth]{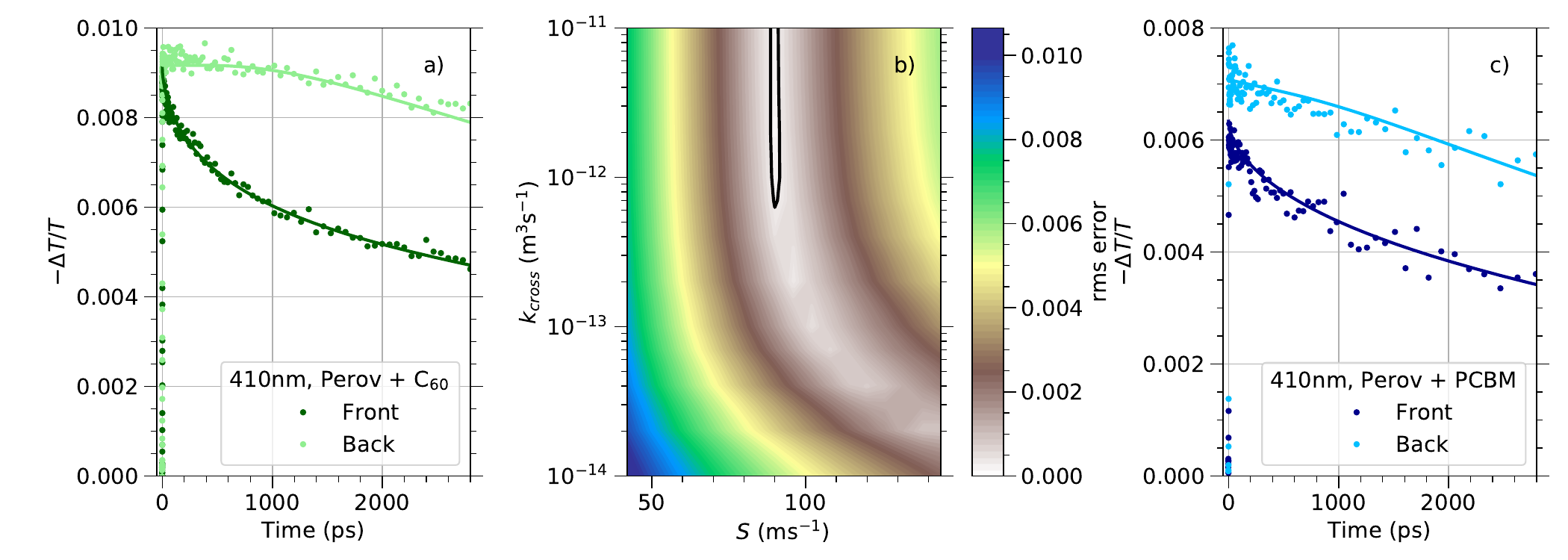}
    \caption{(a) Experimental OPTP results for perovskite-$C_{60}$ bilayer (points) under front and back excitation at 410\,nm, along with fits from the analytical model (solid lines) with $S=90$\,ms$^{-1}$, $\mu=15$\,cm$^2$V$^{-1}$s$^{-1}$ and $\tau_B=200$\,ns.
    (b) Root-mean-square (rms) error determined from the residuals between the numerical simulation and the analytical fit to the data in a), for $\mu=15$\,cm$^2$V$^{-1}$s$^{-1}$ and different values of $S$ and $k_{\mathrm{inter}}$. The black contour illustrates the global minimum error, which occurs at $S=90$\,ms$^{-1}$ and $k_{\mathrm{inter}}\geq 10^{-12}$\,s$^{-1}$.
    (c) Experimental results for perovskite-PCBM bilayers (points) under front and back excitation, along with fits from the analytical model (solid lines) with $S=55$\,ms$^{-1}$, $\mu=15$\,cm$^2$V$^{-1}$s$^{-1}$ and $\tau_B=15$\,ns.
    }
    \label{fig:6-Fitting}
\end{figure*}

Fig.\,\ref{fig:6-Fitting} shows that the global fits of the analytical model (solid lines) accurately reproduce all the features of the OPTP experiments (points), with small residual error (a map of the residual versus $\mu$ and $S$ for the C$_{60}$ bilayer contains a single minimum, as shown in SM Fig.\ S23).
The experimental measurement for the C$_{60}$ bilayer with 410\,nm excitation showed a clear delay in the decay of the signal when exciting on the back (quartz) side, which was captured by the model. 
The values of the best fit parameters were an ambipolar mobility $\mu=15\pm0.6$\,cm$^2$V$^{-1}$s$^{-1}$ (or ambipolar $D=[3.8 \pm 0.2]\times 10^{-5}$\,m$^2$s$^{-1}$) and $S=90\pm2$\,ms$^{-1}$.

Returning to the more detailed two-step numerical model, for the C$_{60}$ bilayers we performed a series of simulations for different values of $S$ and $k_{\mathrm{inter}}$ in order to estimate the values of the two rate constants.
In Fig.\,\ref{fig:6-Fitting}b the rms error between the data reported in Fig.\,\ref{fig:6-Fitting}a (the fitted analytical solution) and the numerical model is shown, as a function of both extraction velocity $S$ and the cross-interfacial recombination rate $k_{\mathrm{inter}}$. 
The numerical simulation best agrees when the rms error is close to zero (white area), inside the area defined by the black contour, corresponding to the region around $S=90$\,ms$^{-1}$ and $k_{\mathrm{inter}} \geq 7\times 10^{-13}$\,m$^3$s$^{-1}$.
The contour plot shows a single minimum, but the contours extend towards large values of $k_{inter}$, indicating this is a lower bound for $k_{inter}$.
The surface recombination velocity in the one step model is equivalent to the extraction velocity in the two step model (both are 90\,ms$^{-1}$).

For the PCBM bilayers, whilst the decay under back excitation was slower compared to excitation on the front (PCBM) side, it did not show a distinct delay to the start of the decay. 
This suggests the solution-deposited PCBM had penetrated into the perovskite's grain boundaries, and that the planar bilayer model may oversimplify a more complex scenario for the PCBM bilayer. 
Alternatively, the solution-phase processing to add the PCBM may have modified the strain and/or defect density in the perovskite, changing its $\tau_B$.
However, a global fit to measurements on PCBM yielded reasonable fits (Fig.\,\ref{fig:6-Fitting}c), with best fit parameters $\mu=15\pm 1$\,cm$^2$V$^{-1}$s$^{-1}$, $S=55\pm2$\,ms$^{-1}$ and $\tau_B=15\pm 2$\,ns.
Therefore we conclude that the electron extraction velocities for PC$_{60}$BM is $S\simeq55$\,ms$^{-1}$, slightly smaller than $S=90\pm2$\,ms$^{-1}$ for the C$_{60}$ bilayer.
We note that Krogmeier \textit{et al.}\ \cite{Krogmeier2018} deduced a value of $S=53$\,ms$^{-1}$ for a perovskite-PC$_{61}$BM bilayer from the power dependence of their PL dynamics on nanosecond timescales, very similar to our values.

\subsection{Further discussion}
By considering the results obtained from the numerical model including the Coulombic forces and interfacial recombination, we demonstrated that the accumulation of electrons in the fullerene layer is undermined by process such as cross-interface recombination or hole extraction. 
This corroborates the interpretation of the SS-PL measurements (Fig.\,\ref{fig:2-SEM PL}e) in which PL quenching by the fullerenes is interpreted as resulting from the introduction of additional channels for non-radiative recombination \cite{Sarritzu2017, Stolterfoht2018, Hutter2020}. 
Cross-interface recombination and hole extraction are difficult to distinguish, as both lead to the loss of a hole from the perovskite and reduce the negative charge in the ETL (they are compared in SM section 2.5). 
Both mechanisms have been discussed in the literature.
Schulz et al.\ \cite{Schulz2015} suggested that the observed tail of states above the C$_{60}$ valence band (VB), reduces the energy offset between the C$_{60}$ and the perovskite VBs and thus may allow some hole extraction into the ETL. 
In contrast, Warby et al.\ \cite{Warby2022} concluded that cross-interface recombination is dominant. 
They observed that adding a C$_{60}$ layer increased the sub-gap feature in the external quantum efficiency (EQE), indicating the presence of intragap states. 
They did not observe any PL or electroluminescence from the C$_{60}$ layer, which could be expected following hole extraction and subsequent radiative recombination within the C$_{60}$. 
Therefore, they proposed recombination is mostly cross-interface via intragap states or broadening of the C$_{60}$ LUMO states due to disorder, which was previously reported \cite{Shao2016}. 
The attribution to cross-interface recombination is further supported by the observation that adding a LiF interlayer a few nanometres thick between the perovskite and C$_{60}$ can reduce the sub-gap feature in EQE (i.e.\ lower the intragap state density) and improve $V_{oc}$. Liu et al.\ \cite{Liu2022} also showed that nanometre thick interlayers improved the performance. Their density functional theory (DFT) simulations showed that when C$_{60}$ is close to the perovskite surface, new states form in the bandgap, but this does not occur if the C$_{60}$ is moved away by an interlayer. 

Finally, regarding the Spiro-OMeTAD-perovskite interface, Eq.\,\ref{Eq:TauS} can be used to roughly estimate the surface loss or removal velocity from the 
PL lifetime. 
The PL data in Fig.\ \ref{fig:2-SEM PL}g shows that the effective lifetime of the Spiro-OMeTAD bilayer ($\approx 100$\,ns) is roughly half that of the perovskite bulk (estimated from the bare perovskite reference), meaning that the surface lifetime, is comparable to the bulk lifetime, $\tau_s \approx \tau_B \approx 200$\,ns. 
Since the second term (i.e.\ the diffusion contribution) in the surface lifetime (Eq.\ \ref{Eq:TauS}) is only $\approx$ 3\,ns, the surface lifetime is dominated by the first term $\tau_s \simeq L/S$. 
$S$ is then estimated as $\approx$ 2ms$^{-1}$, more than 
an order of magnitude below the surface extraction rate for the ETL bilayers.
With no rapid extraction for the HTL, it was not possible to determine the cross-interfacial recombination rate from the comparison between ultrafast spectroscopy and numerical simulation. 

\section{Conclusion}

The combination of picosecond optical spectroscopy and an advanced carrier dynamics model that includes Coulombic forces allowed us to provide an improved understanding of the charge carrier dynamics in perovskite-CTL bilayer heterostructures that employ three of the most common CTLs, namely C$_{60}$, PCBM and Spiro-OMeTAD. 
For perovskite-Spiro-OMeTAD bilayers, negligible decay of the carrier density in the perovskite was observed within the first 3\,ns after photoexcitation, indicating that hole extraction was relatively slow (around 5 to 10 times slower than the electron extraction velocity of the fullerenes). 
Importantly, surface recombination at the interface must also be slow, which may explain the excellent performance of Spiro-OMeTAD in solar cell devices despite its relatively slow hole extraction.
For perovskite-fullerene bilayers, when charge carriers were photoexcited near the fullerene interface, the carrier density decayed within the first few hundred picoseconds and continued to decay on nanosecond timescales. 
The numerical model including Coulombic forces indicated that at the higher carrier densities used in pump-probe experiments, the population decay should plateau within the first nanosecond due to the accumulation of extracted electrons in the fullerene repelling electrons in the perovskite to form an electrical double layer. 
The absence of this characteristic fingerprint of a Coulombic bottleneck indicates that there is substantial loss of holes at the interface.
We highlighted cross-interface recombination and hole extraction as possible physical mechanisms active at the interface, but note that these cannot be distinguished by these experiments.

The numerical and analytical carrier dynamics models presented were found to accurately replicate the experimental data, and determined the interfacial extraction velocity, along with the ambipolar diffusion coefficient for vertical transport through the perovskite films. 
Perovskite-fullerene bilayers were found to fall under the limiting case of rapid interfacial recombination (for which the analytical ``one-step'' model is suitable), and thus only a lower bound for the cross-interface recombination rate constant can be determined.

The trends observed with sub-nanosecond THz and transient absorption measurements were also observed for nanosecond photoluminescence measurements that use lower injected carrier densities which are comparable to solar illumination, indicating that the properties extracted from the sub-nanosecond measurements using higher injected carrier densities are relevant to operational carrier densities.

Having identified targets for improvement in three of the most popular CTLs, future work will build upon the combination of picosecond optical spectroscopy and drift-diffusion modelling presented here, for instance to study the interface extraction and recombination kinetics of other heterojunctions, or to fully elucidate the interfacial recombination processes active at such interfaces.

\appendix
\section{Material synthesis}
In this work, thin films of  $($FA$_{0.83}$MA$_{0.17)})_{0.95}$Cs$_{0.05}$Pb(I$_{0.83}$Br$_{0.17}$)$_3$ were investigated.
All processing was carried out in a N$_2$ glove box with both O$_2$ and H$_2$O levels below 0.1\,ppm. 
Similar results were obtained on films of $($FA$_{0.83}$MA$_{0.17)})_{0.95}$Cs$_{0.05}$PbI$_{3}$. 
Perovskite precursors were prepared such that the (FA + MA): Pb molar ratio was 1:1.1 for both compositions. 
For the mixed I-Br composition, PbI$_2$ (1.1\,mmol, TCI) was added to a DMF:DMSO mix (1\,ml, 4:1 ratio by volume), followed by FAI (1\,mmol, Greatcell solar), PbBr$_2$ (0.22\,mmol, TCI), and then MABr (0.2\,mmol, Greatcell Solar). 
A CsI solution was prepared by dissolving CsI (78\,mg, Sigma Aldrich) in DMSO (200\,$\mu$l). This solution (42\,$\mu$l) was added to the PbI$_2$ + FAI + PbBr$_2$ + MABr in DMF/DMSO mix. 

For the Spiro-OMeTAD layer, Spiro-OMeTAD (72.3 mg, Borun) was dissolved in chlorobenzene (abbreviated CB). 
Separately, a solution of FK209 in acetonitrile (300\,mg ml$^{-1}$) and a solution of Li-TFSi in acetonitrile (520\,mg ml$^{-1}$) were prepared. 
To the Spiro solution, tBP solution (28.8\,$\mu$l), Li-TFSi solution (17.5\,$\mu$l) and FK209 solution (29\,$\mu$l) were added.
For PC$_{60}$BM, powder (20\,mg, Nano-C) was dissolved in CB (1\,ml). 

To make thinner perovskite films, the perovskite solution (100 $\mu$l) was added to the DMF:DMSO 4:1 mix (500\,$\mu$l). 
For the thinner CTLs, the Spiro-OMeTAD (50\,$\mu$l) and PC$_{60}$BM solutions were added to CB (150\,$\mu$l).

The perovskite layers were formed by spin-coating on quartz substrates using an anti-solvent method. 
The perovskite solution was dropped onto the substrate prior to spinning. 
The spinning was divided into two steps. 
First, the substrate was accelerated up to 1000\,rpm at a rate of 200\,rpm/s, then accelerated after 8\,s to 6000\,rpm at a rate of 1000\,rpm/s. 150 $\mu$l of CB was dropped onto the spinning substrate 30\,s into the second step. 
The sample was then annealed at 100\,$^{\degree}$C for 1 hour.

To form the Spiro-OMeTAD layer, the perovskite sample was accelerated to 4000rpm at a rate of 2000rpm/sec, and the Spiro-OMeTAD solution was dropped onto the spinning perovskite sample 5 seconds after starting (spinning stopped at 25 seconds). 
To form the PC$_{60}$BM layer, the PC$_{60}$BM solution was dropped onto the stationary perovskite sample, which was then accelerated to 1000rpm at a rate of 1000rpm/sec (spinning stopped at 30 seconds). 
The C$_{60}$ layers were deposited by thermal evaporation at a rate of 0.1-0.4\,\r{A}s$^{-1}$.

Top down SEM images were recorded with a Zeiss SUPRA 55-VP, and cross-sectional images were taken with a Zeiss Gemini 500.
        
\section{Steady-state optical and time-resolved PL spectroscopy}
UV-visible spectroscopy was performed using a spectrophotometer (Perkin Elmer Lambda 1050) with an integrating sphere module, to determine the film's absorbance taking account of scatter and reflection losses.
The steady state (SS) and time resolved (TR) PL were measured using a Horiba Jobin Yvon Fluorolog-3 fluorescence spectrometer, comprising a iHR320 monochromator and a Hammamatsu R982P PMT.
TR-PL was measured by time-correlated single photon counting (TCSPC), with the timing electronics provided by a Horiba FluoroHub. 
The samples were illuminated with pulsed diode sources with centre wavelengths of 405\,nm and 633\,nm, pulse energies of $\sim$3.4 and $\sim$1.7 pJ/pulse respectively, and pulse duration $<200$\,ps. 
The angle of incidence was $\sim 45^{\circ}$ and the Gaussian illumination spots yielded maximum pulse fluences of 25 and 10nJcm${-2}$ per pulse ($ 5\times 10^{14}$ and $3\times 10^{14}$ photons/m$^2$ per pulse) for 405\,nm and 633\,nm excitation, respectively. 
The PL was collected at $45^{\circ}$. 
A long pass filter was placed between the sample and the spectrometer, and the spectrometer was set to detect at the peak of the PL spectrum (760\,nm).
The fraction of photoexcitation cycles that resulted in detection of a photon (called count rate) was kept below 2\%. 
The PL decays were much longer than the instrument response function (IRF, 2ns), and so iterative reconvolution of the data and the IRF was not necessary. 
The repetition period of the pulsed diodes was chosen to ensure the PL intensity decayed to zero before the next pump pulse arrived (typically 1-10\,$\mu$s). 
To normalise the TCSPC decays, the background offset was first subtracted, and then the data was normalised to a value just after the initial small spike resulting from residual scatter of the pump light.

The relatively low fluence resulted in a low carrier density injection into the perovskite, and carrier recombination was therefore dominated by defect-mediated recombination \cite{Johnston2016} which is thus sensitive to changes in the defect density due to illumination or atmosphere \cite{Meggiolaro2017,Brenes2018,Quitsch2018,Motti2019}.
The TCSPC decay profile can thus depend on the environmental exposure of the sample and may change during an experimental session.
To study the differences between the CTLs we adopted an experimental protocol that minimised any differences in surface defect density during measurements. 
Samples were measured in ambient air starting after several minutes of exposure, until the PL lifetime had stabilised, so that all samples had the opportunity to reduce their surface defect density to a similar level. 
This resulted in bright, long-lived PL dynamic data that reproduced when repeatedly measured, making differences due to the CTL easier to distinguish. 
Alternative environments (e.g.\ nitrogen) resulted in shorter PL lifetimes which makes differences due to CTLs harder to disinguish.

\section{Ultrafast spectroscopy}
Time-resolved measurements of carrier densities were performed using transient absorption (TA) spectroscopy and optical pump terahertz probe (OPTP) spectroscopy, which measured dynamics with sub-ps resolution and over a 3\,ns range. The OPTP setup is described in detail in Ref.\ \cite{Ren2023}. Photon fluxes per pulse for OPTP and TA were calculated by considering the effective area of the overlapping pump and probe spots. 
For TA, the Gaussian intensity profile of the pump and probe spots had $\sigma \sim 0.4$\,mm and $\sigma \sim 0.04$\,mm respectively. 
For OPTP, the Gaussian profile of the pump and probe spots had $\sigma \sim 1$\,mm and $\sigma \sim 0.4$\,mm respectively (for THz this was the standard deviation of the electric field).

For TA, femtosecond laser pulses were generated by a Spitfire Ace amplified Ti:Sapphire laser system (Spectra Physics), outputting 13\,mJ pulses with a centre wavelength of 800nm at a repetition rate of 1\,kHz. 
A beam splitter directed a fraction of this beam power for use in the TA experiment (another fraction was used for OPTP experiments). 
A second beam splitter directed a fraction of this beam power to an optical parametric amplifier (Light Conversion TOPAS) to generate pump pulses of different wavelengths, and the other fraction of the beam was focussed into a 2\,mm CaF$_2$ window to generate a white light continuum (the window is continuously translated to prevent damage).
A mechanical delay stage was used to vary the path length of the beam that generates the white light pulse, and thus vary the relative time delay between the white light pulse and pump pulse arriving at the sample. 
A mechanical chopper (500\,Hz) was used to alternately block and pass the pump pulse, resulting in the white light pulses measuring pump-on and pump-off conditions alternately.
The white light pulses were directed into a fibre coupled Avantes spectrometer (Avaspec 1650 Fast USB) consisting of a grating and a multi-pixel CCD, to measure the spectra shot-by-shot.
$\Delta$OD spectra were calculated from pairs of pump-on and pump-off spectra.

\section{Theoretical Methods}

The numerical solution to the coupled continuity and Poisson equations was constructed in Python with calculations accelerated by the Numba library.
The coupled equations were solved using the Forward Euler method, and the drift-diffusion equations for current were discretised using the Scharfetter-Gummel scheme to improve stability. \cite{Selberherr1984}
The drift-diffusion expression for the electron charge current is 
\begin{equation}
J_n(x,t) = -q\mu_e n(x,t) \frac{\partial \Phi(x,t)}{\partial x} + q D_e \frac{\partial n(x,t)}{\partial x}    
\end{equation}

\noindent and includes the mobility $\mu_e$, the diffusion coefficient $D_e$, the electric potential $\Phi$ and the electron density $n$. 
A similar expression holds for holes.
The electron and hole diffusion constants were given by $D_{e,h}=\mu_{e,h} k_B T / q$ for mobilities $\mu_{e,h}$.
Here we assumed equal electron and hole mobilities in the perovskite for simplicity.
The mobility of carriers in the CTL is much lower than that of the perovskite, and it was thus assumed that any carriers extracted into the CTL would remain close to the interface within the 3\,ns time window of the experiments, and hence that all CTL carriers contribute to cross-interface recombination. 
Poisson's equation, ${\partial^2 \Phi(x,t)}/{\partial x^2}  = -{\rho_f(x,t)}/{\epsilon_r \epsilon_0 }$, was solved at each time step for free charge density $\rho_f$, where the perovskite layer was assumed to have dielectric constant $\epsilon_r=20$.
Numerical simulations were validated by checking numerical convergence for a variety of discretisation conditions, and by comparison to analytical results in limiting cases (see SM Figs.\,S12-S16 \cite{Note1})
The results presented here used a 3\,nm spatial step, and a 10\,fs time step over the 2.5\,ns simulation window.
The initial photoexcited carrier density and depth profile in the perovskite matched those used in the experiment.

\medskip
\textbf{Supplemental Material} \par 
Supplemental Material associated with this article is available online.

\medskip
\begin{acknowledgements} 
EBC would like to thank the EPSRC (UK) for a PhD studentship. 
RJM would like to acknowledge funding from the EPSRC (EP/V001302/1).
KDGIJ thanks the Equal Opportunities Foundation (Hong Kong) for financial support.
The authors thank Prof.\ Ross Hatton, Stephen York and Dr.\ Eric Hu for assistance with C$_{60}$ film deposition, SEM and AFM studies, respectively.
The authors acknowledge use of the Warwick Centre for Ultrafast Spectroscopy Research Technology Platform.
\end{acknowledgements}

\medskip

%

\end{document}
%